\documentclass[aps,prl,reprint,superscriptaddress,showpacs]{revtex4-1}
\usepackage{graphicx}% Include figure files

\begin{document}

\preprint{}

\title{
Well-developed deformation in $^{42}$Si
}

\author{S.~Takeuchi}
 \email{takesato@riken.jp}
\affiliation{
RIKEN Nishina Center, Wako, Saitama 351-0198, Japan
}
\author{M.~Matsushita}
\altaffiliation[Present address: ]{CNS, University of Tokyo, RIKEN campus, Wako, Saitama 351-0198, Japan }
\affiliation{
RIKEN Nishina Center, Wako, Saitama 351-0198, Japan
}
\affiliation{
Department of Physics, Rikkyo University, Toshima, Tokyo 172-8501, Japan
}
\author{N.~Aoi}
\altaffiliation[Present address: ]{RCNP, Osaka University, Mihogaoka, Ibaraki, Osaka, 567-0047, Japan}
\affiliation{
RIKEN Nishina Center, Wako, Saitama 351-0198, Japan
}
\author{P.~Doornenbal}
\affiliation{
RIKEN Nishina Center, Wako, Saitama 351-0198, Japan
}
\author{K.~Li}
\affiliation{
RIKEN Nishina Center, Wako, Saitama 351-0198, Japan
}
\affiliation{
Peking University, Beijing 100871, People's Republic of China
}
\author{T.~Motobayashi}
\affiliation{
RIKEN Nishina Center, Wako, Saitama 351-0198, Japan
}
\author{H.~Scheit}
\altaffiliation[Present address: ]{Institut f\"ur Kernphysik, Technische Universit\"at Darmstadt, 64289 Darmstadt, Germany}
\affiliation{
RIKEN Nishina Center, Wako, Saitama 351-0198, Japan
}
\author{D.~Steppenbeck}
\altaffiliation[Present address: ]{CNS, University of Tokyo, RIKEN campus, Wako, Saitama 351-0198, Japan }
\affiliation{
RIKEN Nishina Center, Wako, Saitama 351-0198, Japan
}
\author{H.~Wang}
\affiliation{
RIKEN Nishina Center, Wako, Saitama 351-0198, Japan
}
\affiliation{
Peking University, Beijing 100871, People's Republic of China
}
\author{H.~Baba}
\affiliation{
RIKEN Nishina Center, Wako, Saitama 351-0198, Japan
}
\author{D.~Bazin}
\affiliation{
National Superconducting Cyclotron Laboratory, Michigan State University, East Lansing, Michigan 48824, USA
}
\author{L.~C\`aceres}
\affiliation{
Grand Acc\'el\'erateur National d'Ions Lourds, CEA/DSM-CNRS/IN2P3, F-14076 Caen Cedex 5, France
}
\author{H.~Crawford}
\affiliation{
Lawrence Berkeley National Laboratory, Berkeley, California 94720, USA
}
\author{P.~Fallon}
\affiliation{
Lawrence Berkeley National Laboratory, Berkeley, California 94720, USA
}
\author{R.~Gernh\"auser}
\affiliation{
Physik Department, Technische Universit\"at M\"unchen, D-85748 Garching, Germany
}
\author{J.~Gibelin}
\affiliation{
LPC-ENSICAEN, IN2P3-CNRS et Universit\'e de Caen, F-14050, Caen Cedex, France
}
\author{S.~Go}
\affiliation{
Center for Nuclear Study, University of Tokyo, RIKEN campus, 
Wako, Saitama 351-0198, Japan
}
\author{S.~Gr\'evy}
\affiliation{
Grand Acc\'el\'erateur National d'Ions Lourds, CEA/DSM-CNRS/IN2P3, F-14076 Caen Cedex 5, France
}
\author{C.~Hinke}
\affiliation{
Physik Department, Technische Universit\"at M\"unchen, D-85748 Garching, Germany
}
\author{C.~R.~Hoffman}
\affiliation{
Physics Division, Argonne National Laboratory, Argonne, Illinois 60439, USA
}
\author{R.~Hughes}
\affiliation{
Department of Physics, University of Richmond, Richmond, Virginia 23173, USA
}
\author{E.~Ideguchi}
\altaffiliation[Present address: ]{RCNP, Osaka University, Mihogaoka, Ibaraki, Osaka, 567-0047, Japan}
\affiliation{
Center for Nuclear Study, University of Tokyo, RIKEN campus, 
Wako, Saitama 351-0198, Japan
}
\author{D.~Jenkins}
\affiliation{
Physics Department, University of York, Heslington, York YO10 5DD, United Kingdom}
\author{N.~Kobayashi}
\affiliation{
Department of Physics, Tokyo Institute of Technology, Meguro, Tokyo 152-8551, Japan
}
\author{Y.~Kondo}
\affiliation{
Department of Physics, Tokyo Institute of Technology, Meguro, Tokyo 152-8551, Japan
}
\author{R.~Kr\"ucken}
\altaffiliation[Present address: ]{TRIUMF, 4004 Wesbrook Mall, Vancouver, Canada}
\affiliation{
Physik Department, Technische Universit\"at M\"unchen, D-85748 Garching, Germany
}
\author{T.~Le~Bleis}
\altaffiliation[Present address: ]{Physik Department, Technische Universit\"at M\"unchen, D-85748 Garching, Germany}
\affiliation{
Institut f\"ur Kernphysik, Johann-Wolfgang-Goethe-Universit\"at, D-60486 Frankfurt, Germany
}
\affiliation{
GSI Helmholtzzentrum f\"ur Schwerionenforschung D-64291 Darmstadt, Germany
}
\author{J.~Lee}
\affiliation{
RIKEN Nishina Center, Wako, Saitama 351-0198, Japan
}
\author{G.~Lee}
\affiliation{
Department of Physics, Tokyo Institute of Technology, Meguro, Tokyo 152-8551, Japan
}
\author{A.~Matta}
\affiliation{
Institut de Physique Nucl\'eaire, IN2P3-CNRS, Universit\'e de Paris Sud, F-91406 Orsay, France
}
\author{S.~Michimasa}
\affiliation{
Center for Nuclear Study, University of Tokyo, RIKEN campus, 
Wako, Saitama 351-0198, Japan
}
\author{T.~Nakamura}
\affiliation{
Department of Physics, Tokyo Institute of Technology, Meguro, Tokyo 152-8551, Japan
}
\author{S.~Ota}
\affiliation{
Center for Nuclear Study, University of Tokyo, RIKEN campus, 
Wako, Saitama 351-0198, Japan
}
\author{M.~Petri}
\altaffiliation[Present address: ]{Institut f\"ur Kernphysik, Technische Universit\"at Darmstadt, 64289 Darmstadt, Germany}
\affiliation{
Lawrence Berkeley National Laboratory, Berkeley, California 94720, USA
}
\author{T.~Sako}
\affiliation{
Department of Physics, Tokyo Institute of Technology, Meguro, Tokyo 152-8551, Japan
}
\author{H.~Sakurai}
\affiliation{
RIKEN Nishina Center, Wako, Saitama 351-0198, Japan
}
\author{S.~Shimoura}
\affiliation{
Center for Nuclear Study, University of Tokyo, RIKEN campus, 
Wako, Saitama 351-0198, Japan
}
\author{K.~Steiger}
\affiliation{
Physik Department, Technische Universit\"at M\"unchen, D-85748 Garching, Germany
}
\author{K.~Takahashi}
\affiliation{
Department of Physics, Tokyo Institute of Technology, Meguro, Tokyo 152-8551, Japan
}
\author{M.~Takechi}
\altaffiliation[Present address: ]{ExtreMe Matter Institute EMMI and Research Division, GSI Helmholtzzentrum, 64291 Darmstadt, Germany}
\affiliation{
RIKEN Nishina Center, Wako, Saitama 351-0198, Japan
}
\author{Y.~Togano}
\altaffiliation[Present address: ]{ExtreMe Matter Institute EMMI and Research Division, GSI Helmholtzzentrum, 64291 Darmstadt, Germany}
\affiliation{
RIKEN Nishina Center, Wako, Saitama 351-0198, Japan
}
\author{R.~Winkler}
\altaffiliation[Present address: ]{Los Alamos National Laboratory, Los Alamos, NM 87545, USA}
\affiliation{
National Superconducting Cyclotron Laboratory, Michigan State University, East Lansing, Michigan 48824, USA
}
\author{K.~Yoneda}
\affiliation{
RIKEN Nishina Center, Wako, Saitama 351-0198, Japan
}

\date{\today}

\begin{abstract}
Excited states in $^{38,40,42}$Si nuclei have been 
studied via in-beam $\gamma$-ray spectroscopy with 
multi-nucleon removal reactions.
Intense radioactive beams of $^{40}$S and $^{44}$S provided 
at the new facility of the RIKEN Radioactive Isotope Beam Factory 
enabled $\gamma$-$\gamma$ coincidence measurements. 
A prominent $\gamma$ line observed with an energy of 742(8) keV in $^{42}$Si 
confirms the $2^+$ state reported in an earlier study. 
Among the $\gamma$ lines observed in coincidence with the $2^+ \rightarrow 0^+$ transition, 
the most probable candidate for the transition from 
the yrast $4^+$ state was identified, leading to a $4^+_1$ energy of 2173(14) keV.
The energy ratio of 2.93(5) between the $2^+_1$ and $4^+_1$ states 
indicates well-developed deformation in $^{42}$Si at $N=28$ and $Z=14$.
Also for $^{38,40}$Si energy ratios with values of 2.09(5) and 2.56(5) were obtained.
Together with the ratio for $^{42}$Si, 
the results show 
a rapid deformation development of Si isotopes from $N=24$ to $N=28$.
\end{abstract}

\pacs{25.60.-t, 23.20.Lv, 27.40.+z, 29.38.Db}

\maketitle

Shell closures and collectivity 
are important properties that characterize the atomic nucleus. 
Interchange of their dominance along isotopic or isotonic chains 
has attracted much attention. 
The recent extension of the research frontier to nuclei 
far away from the valley of stability 
has revealed several new phenomena 
for neutron- or proton-number dependent nuclear structure.
For example, 
a weakening or even disappearance of shell closures 
occur in several neutron-rich nuclei 
at $N=8$~\cite{NSR2000NA23,NSR2000IW02,NSR2000IW03} 
and $N=20$~\cite{NSR1979DE02,NSR1984GU19,NSR1995MO16}. 
A well known example in the case of $N=20$ is the so-called 
`island of inversion'~\cite{NSR1990WA02} 
located around the neutron-rich nucleus $^{32}$Mg.
The low excitation energy of the first $2^+$ state $E_x(2^+_1)$ 
and large E$2$ transition probability~\cite{NSR1979DE02,NSR1984GU19,NSR1995MO16} 
clearly indicate shell quenching in $^{32}$Mg 
despite the fact that $N=20$ is traditionally a magic number.
The next magic number, $N=28$, 
which appears due to the $f_{7/2}$-$f_{5/2}$ spin-orbit splitting, 
has also been explored~\cite{NSR1993SO06,NSR1996SC31,NSR1997GL02,NSR2004GR20,NSR2006CA26,NSR2007CA35}.   
Weakening of the shell closure 
is seen by the decrease of the $2^+_1$ energy for $N=28$ isotones  
from 3.83 MeV in the doubly-closed nucleus $^{48}$Ca to 1.33 MeV in $^{44}$S
as the proton number decreases from $Z=20$ to $Z=16$.
Another region of shell stability 
has been shown to exist at $N,Z=14$ 
due to the $d_{5/2}$-$d_{3/2}$ spin-orbit splitting. 
The silicon isotopes from $^{28}$Si to $^{34}$Si 
exhibit relatively high 2$^+_1$ energies 
ranging from 1.78 MeV ($^{28}$Si) to 3.33 MeV ($^{34}$Si) 
reflecting the $Z=14$ sub-shell closure.
However, the $2^+_1$ energy gradually decreases 
from $^{36}$Si to $^{40}$Si~\cite{NSR1998IB01,NSR2006CA26}, 
suggesting a development of quadrupole collectivity for isotopes with $N>20$. 

With proton number $Z=14$ and neutron number $N=28$, 
the nuclear structure of $^{42}$Si 
is of special interest. 
A simple but important question that arises is whether 
the weakening of the $N=28$ shell closure continues, 
causing an enhancement of nuclear collectivity, 
or if shell stability is restored owing to a possible doubly magic structure.
A study on $^{42}$Si 
was made by a two-proton removal reaction experiment 
with radioactive $^{44}$S beams at the NSCL~\cite{NSR2005FR19,*NSR2006FR13}.
The small two-proton removal cross section 
was interpreted as evidence for a large $Z=14$ sub-shell gap at $N=28$, 
indicating a nearly spherical shape and 
a doubly closed-shell structure for $^{42}$Si.
Contrary to this result, 
a disappearance of the $N=28$ spherical shell closure around $^{42}$Si 
was concluded by another experimental study at GANIL 
with the same reaction~\cite{NSR2007BA47} 
owing to the observation of a low-energy $\gamma$ line interpreted as 
the transition from the $2^{+}_1$ state at 
770(19) keV to the ground $0^{+}$ state ($0^+_{g.s.}$).
This low $E_x(2^+_1)$ value supports the non-magic nature of $^{42}$Si 
expected from comparison of the $\beta$-decay half-life of $^{42}$Si 
with QRPA calculations~\cite{NSR2004GR20}.

The disappearance of the $N=28$ shell closure for $^{42}$Si 
was theoretically pointed out in several recent studies 
with shell-model and mean-field 
approaches~\cite{NSR2008OT04,*Utsuno:arXiv1201.4077,NSR2009NO01,NSR2011LI47}.
These studies 
predict a well-developed large deformation of the ground and low-lying states. 
For further understanding of the structure of $^{42}$Si 
and, more generally, the mechanism of interchange between the shell closures 
and quadrupole collectivity along the $N=28$ and $Z=14$ chains, 
further experimental input is necessary.
In addition to the energy of the $2^+_1$ state, 
the location of the $4^+_1$ state is of crucial importance, 
since the energy ratio between the $4^+_1$ and $2^+_1$ states ($R_{4/2}$) 
represents the character of quadrupole collectivity: 
a ratio of 2 is expected for harmonic vibration and 
3.33 for rigid-body rotation, as extremes. 
Hence, the systematic study of $E_x(2^+_1)$ and $E_x(4^+_1)$ values 
is useful to deepen our understanding on the mechanism 
for the evolution of nuclear structure in the vicinity of $^{42}$Si. 

The purpose of the present study 
is to find the $4^+_1$ state in $^{42}$Si 
as well as to revisit the $2^+_1 \rightarrow 0^+_{g.s.}$ transition.  
Owing to the secondary $^{44}$S beam with the world-highest intensity 
provided by the RIKEN Radioactive Isotope Beam Factory (RIBF), 
population of the 4$^+_1$ state in $^{42}$Si 
by the two-proton removal reaction was possible, 
and even a $\gamma$-$\gamma$ coincidence analysis 
was enabled to establish the level scheme with the help of 
the high-efficiency detector array 
DALI2 (Detector Array for Low-Intensity radiation 2)~\cite{Takeuchi2003,NSR2009TA08}.
Additionally, 
we studied multi-nucleon removal reactions 
of $^{44}$S and $^{40}$S to populate 4$^+$ states 
in $^{40}$Si and $^{38}$Si 
in order to obtain valuable information on the evolution of 
the quadrupole collectivity in neutron-rich Si isotopes 
provided by the systematic trend of the ratio $R_{4/2}$. 

The experiment was performed at the RIBF operated by RIKEN Nishina Center and 
the Center for Nuclear Study, University of Tokyo.
A primary $^{48}$Ca beam at 345 MeV/nucleon with an average intensity of 70 pnA 
bombarded a 15-mm-thick rotating beryllium target. 
A secondary $^{44}$S or $^{40}$S beam was produced by projectile fragmentation 
and analyzed by the BigRIPS fragment separator~\cite{Kubo200397} as 
in earlier experiments~\cite{NSR2009DO10,Kobayashi:arXiv1111.7196}. 
The energy and intensity of the secondary $^{44}$S ($^{40}$S) beam 
was approximately 210 MeV/nucleon (210 MeV/nucleon) and 
$4\times10^{4}$ particles per second (pps) ($6\times10^{4}$ pps), 
respectively.

A carbon target with a thickness of 2.54 g/cm$^2$ was 
located at the F8 focus for secondary reactions. 
Reaction products were
analyzed using the ZeroDegree spectrometer~\cite{Mizoi2005}, and 
identified using the energy loss ($\Delta E$), 
magnetic rigidity ($B\rho$), and 
time-of-flight (TOF) information.
The $B\rho$ value was obtained from the position at the dispersive focus F9 
measured by parallel plate avalanche counters (PPACs)~\cite{Kumagai2001562}.   
$\Delta E$ was measured by an ionization chamber 
at the achromatic focus F11 and the TOF information was obtained from 
the time difference between plastic scintillators at F8 and F11.
The inclusive cross section for the C($^{44}$S,$^{42}$Si) reaction 
at 210 MeV/nucleon 
was obtained to be 0.15(2) mb, 
which indicates a monotonic rise of the cross section as compared with 
0.12(2) mb at 98.6 MeV/nucleon~\cite{NSR2005FR19,*NSR2006FR13} and 
0.08(1) mb at 39 MeV/nucleon~\cite{NSR2007BA47}.

De-excitation $\gamma$ rays were detected
by the DALI2 array in coincidence with the outgoing $^{42}$Si, $^{40}$Si, and $^{38}$Si particles.  
DALI2 consists of 186 NaI(Tl) detectors surrounding the reaction target that 
cover an angular range of $11^{\circ}-160^{\circ}$ 
with respect to the beam axis.
Typical photo-peak efficiency and energy resolution are 
20\% and 10\% (FWHM), respectively, 
for 1 MeV $\gamma$ rays emitted from 
nuclei moving with $\beta(=v/c) \simeq 0.6$. 
These values were obtained  
by Monte-Carlo simulations using the {\small GEANT4} code~\cite{geant4}.

%%%%%%%%%%%%%%%%%%%%%%%%%%%%%%%%%%%%%%%%%%%%%%%%%%%%%%%%%%%%%%%%%%%%%%%%%%
\begin{figure}
\begin{center}
\includegraphics[width=8.6cm]{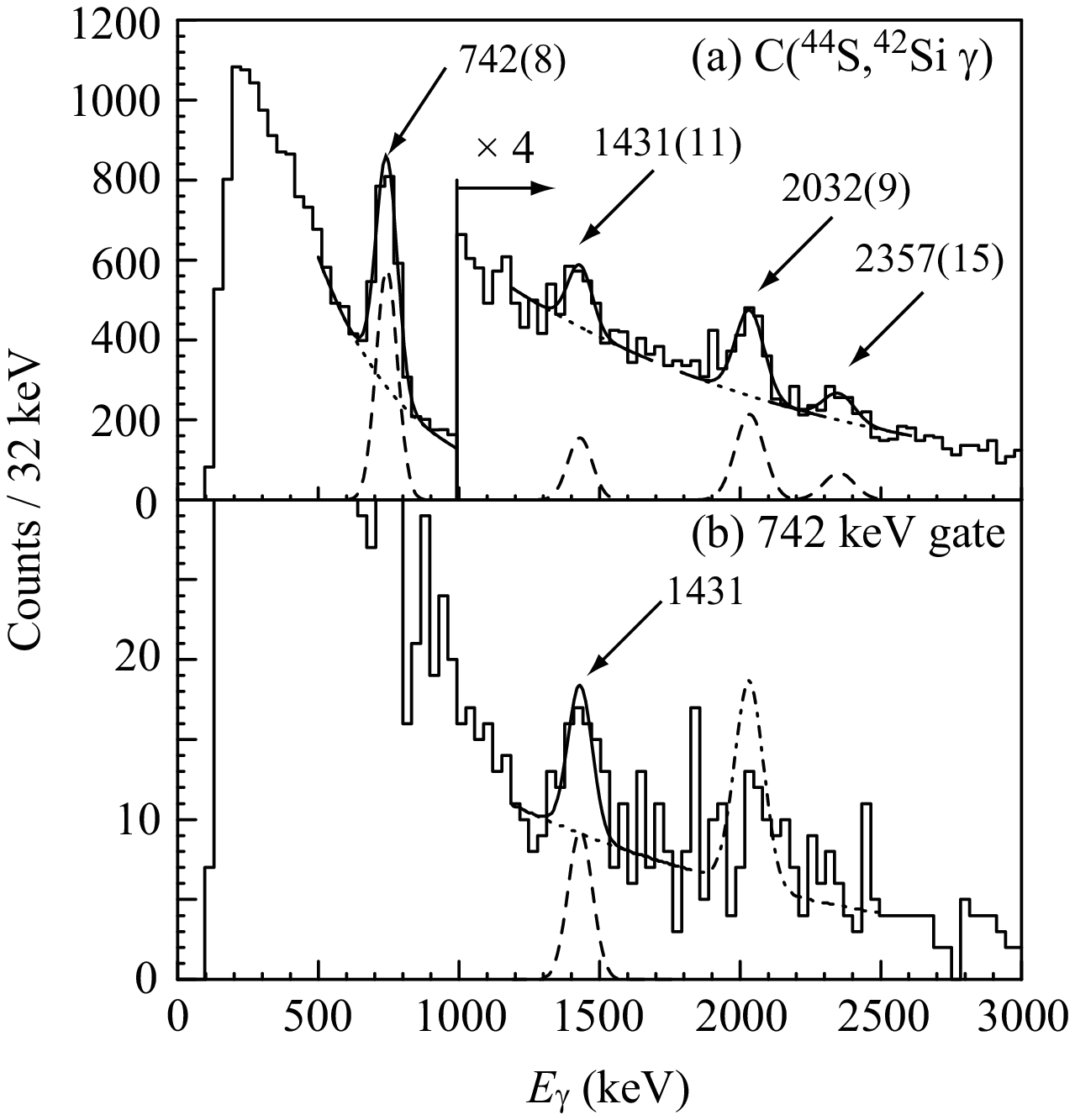}
\caption{\label{fig1}
(a) Doppler-shift corrected $\gamma$-ray spectrum obtained
for the C($^{44}$S,$^{42}$Si $\gamma$) reaction.
(b) $\gamma$-$\gamma$ coincidence spectrum gated with 742 keV $\gamma$ ray.
Solid curves in both figures show the results of fits using Gaussian functions
(dashed curves) and an exponential curve as the background (dotted curve) 
by the method $\chi^2$ minimization.
The dot-dashed line indicates the expected line shape in the case 
that the 2032-keV $\gamma$ ray decays solely to the $2^+$ state.
 }
\end{center}
\end{figure}
%%%%%%%%%%%%%%%%%%%%%%%%%%%%%%%%%%%%%%%%%%%%%%%%%%%%%%%%%%%%%%%%%%%%%%%%%%
Figure~\ref{fig1}(a) shows the $\gamma$-ray spectrum for 
the C($^{44}$S,$^{42}$Si $\gamma$) reaction, 
where Doppler-shift effects have been corrected for. 
For each $\gamma$-ray peak, the energy was obtained 
by a fit of a Gaussian function and an exponential background curve, 
where the energy resolution was fixed to the simulated value. 
As seen in the figure, 
a predominant peak is observed at 742(8) keV, 
where the error includes statistical and systematic uncertainties.  
The systematic error is attributed to the uncertainties 
in the energy calibration (3 keV), 
the Doppler-shift correction (3 keV), 
and 
the ambiguity 
caused by a delay of $\gamma$-emission resulting in a shift of the source position  
(estimated to be 2 keV as a maximum lifetime of $\sim 40$ ps based on systematics ~\cite{NSR2001RA27}). 
The observed peak energy agrees within 1.5 standard deviation with 
the value of 770(19) keV reported in the study at GANIL~\cite{NSR2007BA47}.
In the higher energy region, 
three additional $\gamma$ lines have been identified in the present study 
at 1431(11), 2032(9), and 2357(15) keV.

In order to identify the transitions feeding the $2^+_1$ state, 
a $\gamma$-$\gamma$ coincidence analysis was performed.  
The spectrum obtained from a gate on the 742-keV line 
is shown in Fig.~\ref{fig1}(b).
A clear peak is seen at 1431 keV, 
which may feed the $2^+_1$ state directly 
from a higher-lying excited state at 2173(14) keV. 
By considering the $\gamma$-ray detection efficiency, 
the yield of the peak is consistent with a 100\% feeding of the $2^{+}_1$ state.
A peak-like structure around $E_{\gamma} \simeq 2$ MeV
could correspond to the 2032-keV $\gamma$ line 
observed in Fig.~\ref{fig1}(a), 
providing another candidate that populates the $2^+_1$ state. 
However, the yield of the peak in the $\gamma$-$\gamma$ spectrum is 
lower than the expected value (indicated by the dot-dashed line), 
based on the intensity measured in Fig.~\ref{fig1}(a) 
and assuming a decay branch of 100\% to the level at 742 keV. 
This suggests that the 2032-keV $\gamma$ ray does not, 
or at least does not fully, 
populate the $2^+_1$ state.
From the known tendency that yrast states, 
including the 2$^+_1$ and 4$^+_1$ states, 
are preferentially observed with larger cross sections in nucleon removal reactions 
~\cite{NSR2000BE44,NSR2001YO03,NSR2010FA04},  
those two $\gamma$ lines are possible candidates for 
the $4^+_1 \rightarrow 2^+_1$ transition in $^{42}$Si. 
Since the $\gamma$-$\gamma$ coincidence analysis indicates that 
the 1431-keV $\gamma$ ray directly feeds the $2^+_1$ state as discussed above, 
2173 ($742+1431$) keV for the $4^+_1$ energy is more probable among the two possibilities.
Thus, we tentatively assign 
the $4^+_1$ state at 2173(14) keV.
The resultant $R_{4/2}$ value of 2.93(5) for $^{42}$Si 
is rather close to the rigid-rotor limit. 
This contradicts the possibility of a doubly-closed structure 
suggested by the two magic numbers $Z=14$ and $N=28$, 
but supports enhanced quadrupole collectivity in $^{42}$Si expected 
from the measured $2^+_1$ energy~\cite{NSR2007BA47} and 
theoretical calculations~\cite{NSR2008OT04,*Utsuno:arXiv1201.4077,NSR2009NO01,NSR2011LI47}. 
Furthermore, the large $R_{4/2}$ value indicates 
a large static ground state deformation of $^{42}$Si.

%%%%%%%%%%%%%%%%%%%%%%%%%%%%%%%%%%%%%%%%%%%%%%%%%%%%%%%%%%%%%%%%%%%%%%%%%%
\begin{figure}
\begin{center}
\includegraphics[width=8.6cm]{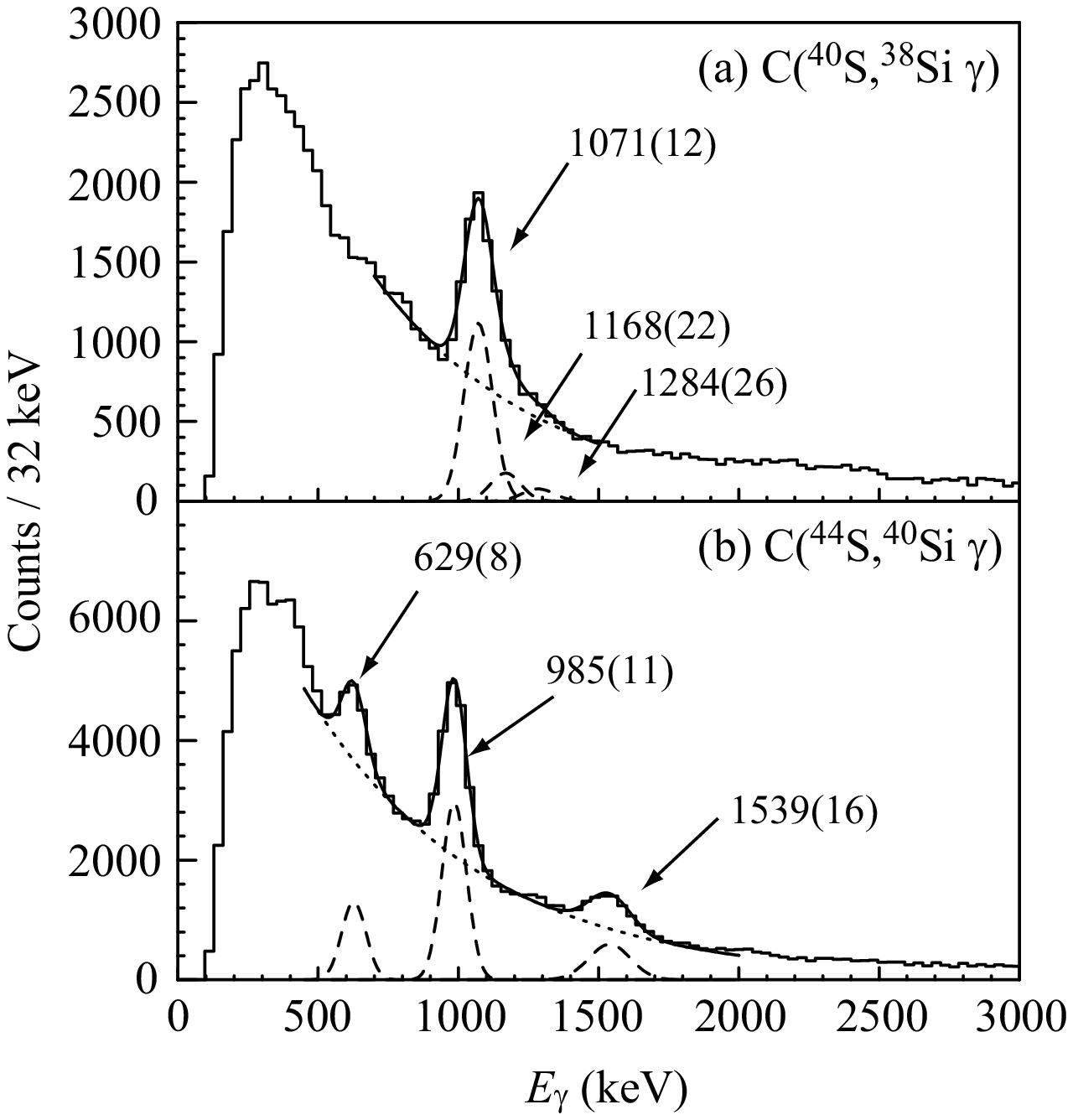}
\caption{\label{fig2}
Doppler-shift corrected $\gamma$-ray spectrum obtained
in (a) the C($^{40}$S,$^{38}$Si $\gamma$) reaction and 
(b) the C($^{44}$S,$^{40}$Si $\gamma$) reaction.
 }
\end{center}
\end{figure}
%%%%%%%%%%%%%%%%%%%%%%%%%%%%%%%%%%%%%%%%%%%%%%%%%%%%%%%%%%%%%%%%%%%%%%%%%%
A search for $4^+_1$ states in $^{38,40}$Si was also conducted.
Figures~\ref{fig2}(a) and (b) show the Doppler-corrected spectra 
for $^{38}$Si and $^{40}$Si, 
measured with the C($^{40}$S,$^{38}$Si $\gamma$) and 
C($^{44}$S,$^{40}$Si $\gamma$) reactions, respectively.
In the case of $^{38}$Si, the major peak likely consists of three unresolved 
$\gamma$-ray lines, according to the ones 
observed in a previous experiment~\cite{NSR2007CA35}. 
Their energies, 1071(12), 1168(22), and 1284(26) keV, 
were obtained by fitting the spectrum 
shown in Fig.~\ref{fig2}(a), 
where three peaks with fixed widths 
and initial centroid positions estimated from Ref.~\cite{NSR2007CA35} 
were used in the fitting procedure. 
The 1071-keV $\gamma$ line corresponds to the known 
$2^+_1 \rightarrow 0^+_{g.s.}$ transition,  
while the 1168- and 1284-keV lines are candidates for 
the $4^+_1 \rightarrow 2^+_1$ $\gamma$ ray.
Here, we assign the 2239-keV state to be the most probable candidate 
for the $4^+_1$ state, based on the same arguments of yrast feeding 
that were discussed for $^{42}$Si.
We note that the alternative $4^+_1$ assignment to the 2355-keV state 
does not affect the discussion 
on the systematics of the level energies of Si isotopes discussed later.

For $^{40}$Si, three $\gamma$ lines 
are seen at 629(8), 985(11), and 1539(16) keV in Fig.~\ref{fig2}(b). 
The 629- and 985-keV lines 
have been observed in the $p$($^{42}$P,$^{40}$Si$+\gamma$) reaction. 
The 985-keV $\gamma$ ray was assigned to the $2^+_1 \rightarrow 0^+_{g.s.}$ transition~\cite{NSR2006CA26}.  
The line at 1539 keV is reported here for the first time. 
According to $\gamma$-$\gamma$ coincidence analysis, 
the 629- and 1539-keV $\gamma$ lines 
are considered to be transitions 
from the excited states at 1614(14) and 2524(19) keV to the $2^+_1$ state. 
Using similar arguments to those given for $^{38}$Si and $^{42}$Si 
regarding the preferential population of yrast states, 
either of the two states could be the yrast $4^+$ level. 
However, systematic trends of $2^+$ and $4^+$ energies suggest that 
the level at 2524 keV is the more likely of the two.
The 1614-keV state is consistent with the previously observed tentative 
1624(7)-keV state~\cite{NSR2006CA26}. 
On the basis of the large-scale shell model (SM) calculations 
in a $\pi(sd)^{Z-8}\nu(pf)^{N-20}$ model space~\cite{NSR2006CA26} 
this level could be either a $0^{+}$ or $2^{+}$ level.
%%%%%%%%%%%%%%%%%%%%%%%%%%%%%%%%%%%%%%%%%%%%%%%%%%%%%%%%%%%%%%%%%%%%%%%%%%
\begin{figure}
\begin{center}
\includegraphics[width=8.6cm]{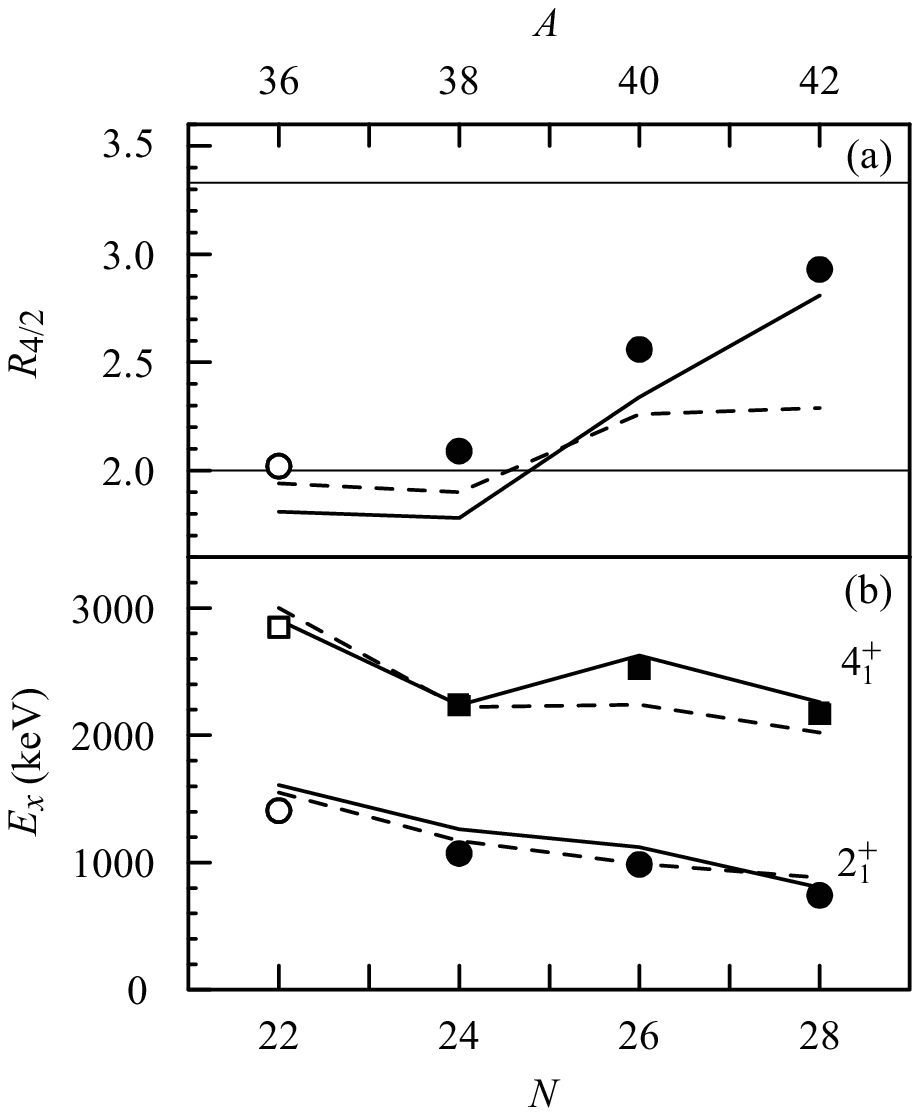}
\caption{\label{fig3}
(a) Ratio between the energies of the $2^+_1$ and $4^+_1$ states ($R_{4/2}$) for Si isotopes. 
The horizontal lines at 2.0 and 3.3 indicate the vibrational and rotational limits, respectively. 
(b) Excitation energies for $2^+_1$ and $4^+_1$ states, 
which are indicated by circles and squares, respectively. 
Filled symbols are results of the present study, 
and solid and dashed lines represent predictions of 
the SM with {\small SDPF-MU}~\cite{NSR2008OT04,*Utsuno:arXiv1201.4077} 
and SM with {\small SDPF-U-MIX} ~\cite{PhysRevLett.109.092503}, respectively (see text for details).
The $2^+$ energies of the $N=24, 26, 28$ Si isotopes have been 
measured in previous works~\cite{NSR1998IB01,NSR2006CA26,NSR2007BA47}.
}
\end{center}
\end{figure}
%%%%%%%%%%%%%%%%%%%%%%%%%%%%%%%%%%%%%%%%%%%%%%%%%%%%%%%%%%%%%%%%%%%%%%%%%%

%%%%%%%%%%%%%%%%%%%%%%%%%%%%%%%%%%%%%%%%%%%%%%%%%%%%%%%%%%%%%%%%%%%%%%%%%%%%
\begin{figure}
\begin{center}
\includegraphics[width=8.6cm]{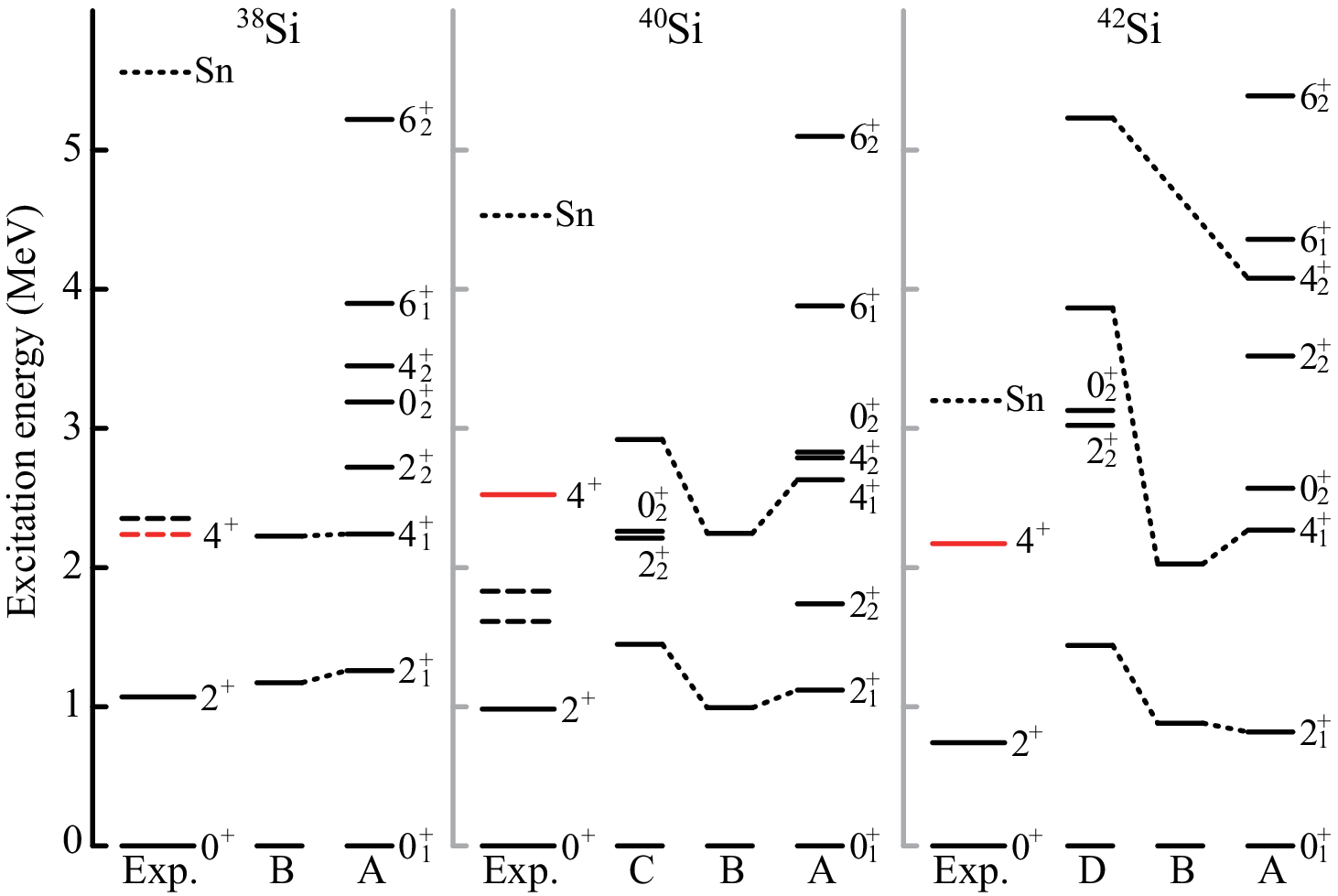}
\caption{\label{fig4}
(Color online)
Summary of the experimental and theoretical levels 
in $^{38,40,42}$Si isotopes. 
The dotted lines with the label S$_n$ indicate 
the location of the experimental (for $^{38,40}$Si) or 
evaluated (for $^{42}$Si) neutron separation energy~\cite{NNDC}.
Among the experimental levels, 
the lines with red color indicate the new results from the present study. 
The labels A, B, C, and D represents respectively 
the results of the shell model calculations of 
Refs.~\cite{NSR2008OT04,*Utsuno:arXiv1201.4077,PhysRevLett.109.092503} and \cite{NSR2006CA26}, 
and mean field calculation of Ref.~\cite{NSR2011LI47}.
}
\end{center}
\end{figure}
%%%%%%%%%%%%%%%%%%%%%%%%%%%%%%%%%%%%%%%%%%%%%%%%%%%%%%%%%%%%%%%%%%%%%%%%%%%%

Figure~\ref{fig3} shows the isotopic dependence of 
the excitation energies of the $2^+_1$ and $4^+_1$ states 
together with their ratio $R_{4/2}$ for $^{36 \textendash 42}$Si, 
where filled symbols represent the present results, 
and open symbols for $^{36}$Si are taken from Ref.~\cite{NSR2006LI32}. 
As seen in the figure, 
the ratios for $^{36}$Si and $^{38}$Si [$R_{4/2} =$ 2.024(4) and 2.09(5)] 
are close to the vibrational limit, suggesting a nearly spherical shape, 
whereas $R_{4/2}$ for $^{40}$Si increases to 2.56(5),
indicating a deviation from the spherical shape or enhancement 
of quadrupole collectivity at $N=26$. 
The lowering of the $2^+_1$ energy as well as 
the increase of the reduced transition probability 
or the deformation parameter obtained for $^{36,38,40}$Si was 
interpreted as a narrowing of 
the $N=28$ shell gap~\cite{NSR1998IB01,NSR2006CA26,NSR2007CA35}. 
The prediction of the SM calculation using the recent {\small SDPF-U-MIX} interaction~\cite{PhysRevLett.109.092503}, 
an updated version of {\small SDPF-U}~\cite{NSR2009NO01}, 
is indicated by the dashed lines in Fig.~\ref{fig3}. 
The new interaction, which allows $np$-$nh$ excitations across the $N=20$ shell gap, 
reproduces $E_x(2^+_1)$ and $R_{4/2}$ in a satisfactory manner up to $^{40}$Si, 
but then deviates significantly from the experimental result for $^{42}$Si in Fig.~\ref{fig3}(a).
In the case of the $N=28$ isotope, 
$E_x(2^+_1)$ is lower than $^{40}$Si and $R_{4/2}$ 
further increases to 2.93(5) despite the neutron magic number $N=28$. 
This indicates that a development of nuclear deformation continues up to at least $N=28$.
In addition, the quadrupole collectivity increase in proton deficient $N=28$ isotones 
turns out to continue to $^{42}$Si with $Z=14$. 
Thus, for $^{42}$Si well developed deformation has been experimentally established, 
and the possibility of a doubly-magic nature has been excluded. 

The solid lines in Fig.~\ref{fig3} indicate the results obtained by 
the SM calculation using the {\small SDPF-MU} interaction~\cite{NSR2008OT04,*Utsuno:arXiv1201.4077}, 
which includes the tensor force in the effective interaction. 
The model reproduces the overall trends well, particularly for the $4^+_1$ energies, 
where better agreement is achieved compared to {\small SDPF-U-MIX}. 
In particular, it predicts a large $R_{4/2}$ value of 2.8 for $^{42}$Si 
which is close to the experimental result of 2.93(5).
The mean field calculation with 
the relativistic energy density functional DD-PC1~\cite{NSR2011LI47} 
predicts a rotational band in $^{42}$Si. 
Though the excitation energies of the $2^+_1$ and $4^+_1$ states 
are about two times larger than the experimental values, 
the $R_{4/2}$ ratio (2.7) is not far from the present result. 

Figure~\ref{fig4} summarizes the experimental and theoretical 
level-energies for the isotopes $^{38,40,42}$Si. 
The dashed lines 
indicate the levels observed 
in earlier studies~\cite{NSR2006CA26,NSR2007CA35} 
and the red lines 
show the $4^+$ states tentatively 
assigned in the present study. 
The labels A, B, C, and D represent theoretical calculations by 
the SM with {\small SDPF-MU}~\cite{NSR2008OT04,*Utsuno:arXiv1201.4077}, 
SM with {\small SDPF-U-MIX}~\cite{PhysRevLett.109.092503}, 
SM in Ref.~\cite{NSR2006CA26}, and 
mean field calculation~\cite{NSR2011LI47}, respectively. 
As seen in the figures, some model calculations predict 
more levels compared with the ones experimentally observed. 
The experimental energies of the calculated states with no empirical counterparts 
may provide a deeper understanding of nuclear structure.
Together with possible measurements on $\gamma$-ray angular 
distributions and/or correlations, further efforts to identify those states 
present one target for future experiments with higher statistics by 
improvement of RI beam production. 
Other challenges on the experimental front, 
such as measuring excited states in $^{40}$Mg and $^{44}$Si, 
are encouraged to further trace 
the quadrupole-collectivity development 
along the chains of $N=28$ isotones and $Z=14$ isotopes. 

In summary, 
excited states in $^{38,40,42}$Si 
have been investigated via in-beam $\gamma$-ray spectroscopy 
with multi-nucleon removal reactions in inverse kinematics 
by using 210-MeV/nucleon $^{40,44}$S beams.
With the high-efficiency detector array DALI2 and 
the high intensity secondary beams provided at RIKEN RIBF, 
measurements with high statistics were achieved. 
The energy of the first $2^+$ state in $^{42}$Si 
was measured to be 742(8) keV, 
and the most probable candidate of the $4^+_1$ state 
was found at 2173(14) keV with the aid of 
a $\gamma$-$\gamma$ coincidence analysis. 
The $4^+_1$ states in $^{38}$Si and $^{40}$Si 
were assigned excitation energies of 2239(25) and 2524(19) keV, respectively. 
The systematics of the ratio $R_{4/2}$ of the $4^+_1$ and $2^+_1$ energies 
in silicon isotopes from $N = 24$ to $N = 28$ 
shows a rapid development of deformation. 
The $R_{4/2}$ value of 2.93(5) for $^{42}$Si 
is the largest also among the known $N=28$ isotones, 
indicating that this nucleus has a character of 
well-deformed rotor despite the magic numbers $N = 28$ and $Z = 14$. 

\begin{acknowledgments}
We thank the RIBF accelerator staff and BigRIPS team 
for their operation during the experiment. 
We thank A.~Poves for providing the new values and also Y.~Utsuno and T.~Otsuka 
for their calculations and discussions.
This work was supported by the DFG (EXC 153, KR 2326/2-1), 
the U.S Department of Energy under Contact No. DE-AC02-05CH11231 and DE-AC02-06CH11357, 
and the JUSEIPEN program. 
\end{acknowledgments}

%\bibliography{paper-15}

%merlin.mbs 2010-03-15 4.21a (PWD, AO, DPC)

\end{document}